\documentclass[iop,apj]{emulateapj}
\usepackage{amsmath}
\usepackage{bm}
\usepackage{times}
\newcommand{\Fig}[1]{Figure~\ref{#1}}
\newcommand{\Eq}[1]{Equation~(\ref{#1})}
\newcommand{\EQ}{\begin{equation}}
\newcommand{\EN}{\end{equation}}

\newcommand{\Sec}[1]{Section~\ref{#1}}
\newcommand{\Tab}[1]{Table~\ref{#1}}
\newcommand{\St}{\text{St}}
\newcommand{\er}{\hat{r}}
\newcommand{\MM}{\mbox{\boldmath $M$}{}}
\newcommand{\FF}{\mbox{\boldmath $F$}{}}
\graphicspath{{.}{./Figures/}}

\def\icarus{{Icarus}}           

\date{\today,~ $ $Revision: 1.20 $ $}
\begin{document}

\title{Ferromagnetism and particle collisions: applications to protoplanetary disks and the meteoritical record}
\shorttitle{Ferromagnetism and dust collisions}

\author{Alexander Hubbard\altaffilmark{1}}
\altaffiltext{1}{American Museum of Natural History, New York, NY, USA}
\email{{\tt ahubbard@amnh.org}}

\begin{abstract}
The meteoritical record shows both iron partitioning and tungsten isotopic partitioning between
matrix and chondrules. Tungsten is not abundant enough to have driven its own isotopic partitioning, but if tungsten
were correlated with iron, then ferromagnetic interactions grains could help explain both observations.
We derive a practical parameterization for the increase in particle-particle collision rates
caused by mutually attracting particle magnetic dipole moments.
While the appropriate magnetic parameters remain uncertain, we show that ambient magnetic fields in 
protoplanetary disks are expected to be strong enough to magnetize iron metal bearing dust grains 
sufficiently to drive large increases in their collision rates. Such increased collision rates
between iron metal rich grains could
help preserve primordial iron and W isotopic inhomogeneities; and would help explain
why the meteoritical record shows their partitioning in the solar nebula.
The importance of magnetic interactions
for larger grains whose growth is balanced by fragmentation is less clear, and will
require future laboratory or numerical studies.
\end{abstract}



\section{Introduction}

Laboratory examinations of meteorites provide a fascinating window into dust processing during the first stages
of planet formation. Chondritic meteorites, being undifferentiated,
record the nebular state of the dust with some fidelity \citep{2003ApJ...591.1220L}.
In particular, they are named after
the chondrules they contain: sub-mm melted glassy beads \citep{1997AREPS..25...61H}. While chondrules were clearly
dramatically thermally processed, they nonetheless inform us about their
 collisionally grown precursors.
 
One bizarre feature of chondrites is that chondrules are strongly depleted of iron relative to the inter-chondrule,
non-thermally processed matrix material. Indeed, iron metal abundance is one of the primary
vectors along which meteoritical compositions vary \citep{1982GeCoA..46.1081G}.
There have been many studies of the behavior of iron under chondrule forming temperatures,
and the difficulty in reproducing observations argues strongly for differing chondrules
to have been formed in regions of differing chemical abundances \citep{1996cpd..conf..243G,2004GeCoA..68.1677C,2008M&PS...43.1725E}.
While large variations in rare element abundances might be expected through nugget effects,
iron represents approximately $20\%$ of a chondrite's mass \citep{2003ApJ...591.1220L}.
Chondrules precursors would have been formed by agglomerating billions of
extremely sticky \citep{2010A&A...513A..56G}
sub-micron pre-solar dust grains, and that agglomeration process should have washed out any primordial compositional inhomogeneities.
Nonetheless, to reproduce observations it seems necessary for iron, or some other element abundant enough to react with a significant fraction of the iron,
to have been partitioned between different classes of dust grain. It is unclear how that could have occurred.

\cite{2015LPI....46.2262B} further found tungsten isotopic partitioning between chondrules and matrix. Tungsten is extremely refractory
and isotopic differences do not alter chemistry, making the W isotopic partitioning even more puzzling than that of iron.
Tungsten is also far too rare to affect the process of dust growth itself.
The observed W isotope partitioning requires that tungsten have been carried into the solar nebula by at least two different families of
pre-solar grains with different isotopic abundances. It further requires that the dust coagulation process have kept those
families separate, and finally that the chondrule formation process have treated those families differently.

One way for iron, and tungsten isotopes, to have been partitioned between matrix and chondrules is for matrix and chondrules
to have been produced in different spatial locations, or at different times, from ambient dust of different pre-solar origins.
This possibility is unlikely because of complementarity: the composition of matrix and chondrules are strongly correlated \citep{2010E&PSL.294...85H}.
Further, \cite{2015MNRAS.452.4054G} showed that that correlation requires matrix and chondrules to have been co-genetic. Accordingly, there is a need
for a model of dust growth which takes a well-mixed initial cloud of dust and manages to preserve primordial inhomogeneities
between grains within that cloud during the dust coagulation phase. The model must further
allow for the chondrule formation to treat dust grains of differing compositions differently.

\section{Invoking magnetic interactions}

In this paper, we mostly leave aside the chondrule formation process, which we discuss in a companion paper \citep{Relative_Settling}.
Instead, we focus on
the dust coagulation.
We require a process for preserving dust families with differing primordial compositions
that can operate even at the sub-micron interstellar grain scale \citep{Draine2003}.
Given that tiny interstellar grains are extremely sticky \citep{2010A&A...513A..56G}, this rules out
processes that only effect the surface stickiness: the need to preserve the families starting from interstellar, sub-micron,
completely sticky grains means that the composition of the grains needs to correlate with the collision rate itself, not just the sticking
fraction rate of the collisions. We will also address the issue of how large the difference in the dust coagulation process
needs to be for the families to end up with different aerodynamics. Differing dust aerodynamics allows for dust spatial sorting,
which would allow a spatially restricted chondrule formation mechanism to process the families at different
rates as is discussed in detail in \cite{Relative_Settling}.

In \cite{2014arXiv1407.0274H} we examined how dust-dust collisions were affected by the interaction of
magnetic dipole moments  induced in their iron-metal component by the strong nebular magnetic fields
expected in the solar nebula near Mercury's orbit.
We found that the magnetic interactions were extremely strong, and could erode silicates leaving behind large
iron metal rich dust particles.
Inspired by that work, and by the iron partitioning problem, we here examine how important it is to include the interaction of
magnetic dipoles induced by the much weaker nebular magnetic fields expected in protoplanetary disks
at the orbital position of the asteroid belt. Given the strong uncertainties in many of the parameters, in particular the abundance
and relative magnetic permeability of astrophysical iron-nickel alloys, our primary result is the derivation
a practical parameterization for how effectively magnetic dipole interactions promote dust grain collisions
and collisional growth.

That parameterization suggests that magnetic interactions should indeed have
played a major role in the first stages of dust growth, dramatically enhancing the dust-dust collision rates
and promoting collisions between iron-rich dust grains over collisions between iron-poor grains. That latter 
would have helped preserve primordial iron inhomogeneities.
It also provides a way to preserve primordial tungsten isotopic anomalies, so long as those anomalies 
were correlated with grain ferro-magnetism.
However the magnitude of the effect is much
lower at the dust size scale where collisional growth is balanced by fragmentation.
We suggest that while magnetic attraction would not have strongly altered that equilibrium grain size, it
could have helped prevent iron metal rich dust grain sub-components from fragmenting, allowing magnetic erosion \citep{2014arXiv1407.0274H}
to have proceeded, generating iron metal inhomogeneities. Further, as is discussed in \cite{Relative_Settling},
even modest differences in equilibrium grain size could lead to significant differences in chondrule processing rates.

\section{Interactions between magnetized dust grains}
\label{sec_mag_interactions}

\subsection{Parameter definitions}

We follow the formalism of \cite{2014arXiv1407.0274H}, with 
$m$, $m_i$ and $f \equiv m_i/m$ denoting a dust grain's total mass, iron metal mass,
and metal mass fraction respectively.  Dust grains are assumed
spherical, with radii $a$, density $\rho$ and volume
\EQ
\forall= \frac {4\pi}{3} a^3 = \frac{m}{\rho},
\EN
of which 
\EQ
\forall_i = \frac{m_i}{\rho_i}
\EN
is taken up by metal with density $\rho_i$.
These definitions allow us to calculate the metal volume fraction $g$:
\EQ
g \equiv \frac{\forall_i}{\forall} = \frac{m_i\, \rho}{m\, \rho_i} = f \frac{\rho}{\rho_i}. \label{gdef}
\EN
Note that collisionally aggregated dust grains in protoplanetary disks are expected to be porous.
Accordingly, the densities of iron metal, fused solids, and porous solids, are, respectively \citep{Friedrich2014}:
\begin{align}
&\rho_i \simeq 7.86 \text{\,g\,cm}^{-3}, \\
&\rho_s \simeq 3 \text{\,g\,cm}^{-3}, \\
& \rho_{\phantom{s}} = \phi\, \rho_s, \label{rhophi}
\end{align}
where $\phi$ is the volume filling fraction,

Outside of limited regions of extremely strong magnetic fields \citep{2014arXiv1407.0274H}, the ambient magnetic field $H$
in protoplanetary disks is generally too weak to magnetize iron metal near to saturation.\footnote{
Studies of magnetic fields in protoplanetary disks generally use the $B$ field rather than the $H$ field because
the contribution of magnetized solids is negligible.}
We therefore use the relationship
\EQ
M= \frac{\mu_r H}{4 \pi},
\label{Mmur}
\EN
where $M$ is the magnetization induced by the ambient field and $\mu_r$ is the magnetic relative permeability when
considering magnetically soft materials, which are the main thrust of this paper.
Further, we assume that the iron metal is
evenly distributed throughout our dust grains.
We parameterize the ambient magnetic field $H$ in terms of the thermal pressure through the plasma $\beta$ parameter:
\EQ
\beta \equiv \frac{8 \pi n k_B T}{H^2} = \frac{8 \pi \rho_g k_B T}{\bar{m} H^2},
\EN
where $n$ is the number density of the gas,
$\rho_g$ its density and $\bar{m}$ its the mean molecular mass.
It follows that
\EQ
H^2 = \frac{8 \pi \rho_g k_B T}{\bar{m} \beta}. \label{H_beta}
\EN
Given the strongly radially decreasing gas density expected in protoplanetary disks, \Eq{H_beta} implies
that there should exist an outer radius outside of which induced magnetization is negligible.

\subsection{Parameter normalizations}

\begin{table}
\caption{Magnetic interaction parameters}
\centering
\begin{tabular}{l l l}
\hline \\  [-2ex]
Symbol & Normalization & Parameter \\
\hline \\ [-2ex]
${\mu}_r $ & $1000$ & Metal relative permeability \\
$f$ & $0.2$ & Metal mass fraction \\
${g}$ & $0.0076$ &  Metal volume fraction \\
$\phi$ & $0.1$ & Grain volume filling fraction \\
${\beta}$ & $100$ & Plasma beta \\
${\rho}_g$ & $10^{-11}$\,g\,cm$^{-3}$ & Gas density (MMSN at 2.5 AU) \\
$\Sigma_g$ & $430$\,g\,cm$^{-3}$ & Gas surface density  (MMSN at 2.5 AU)\\
$T$ & $177$\,K & Gas temperature (MMSN at 2.5 AU)\\
$c_s$ & $8 \times 10^4$\,cm\,s$^{-1}$ & Gas sound speed  (MMSN at 2.5 AU)\\
$\bar{m}$ & $2.33$\,amu & Gas mean molecular mass \\
$\bar{C}$ & $0.2$ & Cross-section parameter (Equation~\ref{sigma_m}) \\
\hline
\end{tabular}
\label{normalizations}
\end{table}

We use the gas parameters of a Hayashi MMSN's midplane at $R=2.5$\,AU, appropriate for the asteroid belt \citep{1981PThPS..70...35H}.
Unfortunately, most of the magnetic parameters used in this paper are poorly constrained.  The values we normalize to,
denoted with overbars, are listed in \Tab{normalizations}.  The relative permeability
of iron metal, $\mu_r$, depends on its manufacture, composition, history and the amplitude of the ambient field, potentially varying from $0$
up to tens of thousands for some nickel-iron alloys \citep{1975chcp.book.....W}.  We normalize, possibly optimistically,
to $\bar{\mu}_r = 1000$, and hope that this work will inspire future research to shed light on this parameter.

Similarly, the amplitude of the ambient magnetic field $H$ is unclear,
in no small part because we do not fully understand the behavior of the field in magnetically ``dead'' zones such as
the midplane at $R=2.5$\,AU \citep{1996ApJ...457..355G}.
However, recent work has suggested that the Hall effect can allow significant field
growth even there \citep{2014arXiv1402.7102B}, implying that the term ``dead zone'' may be a misnomer, leading us to
normalize to $\bar{\beta} =100$.
For the parameters in \Tab{normalizations}, and using Equations~(\ref{Mmur}) and (\ref{H_beta}), we find
\EQ
\bar{M} \simeq 10 \text{ emu/cc}^3, \label{M_norm}
\EN
significantly less than the saturated magnetization of iron of \citep{1975chcp.book.....W}:
\EQ
M_S \simeq 1720  \text{ emu/cc}^3. \label{M_sat}
\EN

The appropriate iron metal fraction for dust in protoplanetary disks is unknown, but inspired by the low iron fraction
of astrophysical silicates \citep{Draine2003}
 we adopt $\bar{f} \sim 0.2$, the overall chondritic iron mass fraction
\citep{2003ApJ...591.1220L}. To estimate the parameter $g$ we also need
an estimate for the overall grain density (or equivalently the overall grain volume filling factor). Collisionally
grown grains are expected to be highly porous,
but as the collisional speeds approach ones capable of rearranging
the grains, the grains will be collisionally compacted and the  volume filling fractions will rise
\citep{2012A&A...541A..59S}. We adopt $\bar{\phi}=0.1$, which when
inserted into Equations~(\ref{gdef}) through (\ref{rhophi}) implies that
\EQ
\bar{g} = \bar{f}\, \frac{\bar{\phi}\, \rho_s}{\rho_i} = 0.0076,
\EN
although we will be invoking grains both enriched in metallic iron (higher $f,g$), and depleted in metallic iron (lower $f,g$).

\subsection{Magnetically enhanced collisional cross section: magnetically soft case}

Consider two dust grains $1$ and $2$ close to each other, and embedded in an external magnetic field.
It is easiest to operate in either the limit of either identically sized grains or asymptotically different sized grains with $a_1 \gg a_2$.
The theory of turbulently stirred dust collisions is far simpler in the latter case, so we will henceforth require $a_1 \gg a_2$.
Accordingly, we place our larger grain $1$ at the origin and consider it fixed, while following the motion of the smaller grain $2$.
We will assume that the dust grains are perfectly magnetically soft: i.e.~that their 
magnetizations are controlled purely by a spatially uniform external magnetic field $H$ according to \Eq{Mmur};
and we adopt a spherical coordinate
system with $\theta$ the polar coordinate and the axis aligned with the external field (and hence the magnetic dipole moments).
For simplicity we assume that non-size dependent parameters other than $f$ and $g$,
i.e.~$\rho$, $\phi$, $\mu_r$, and $M$, are identical for both grains.

The magnetic energy of the dipole-dipole interaction is
\EQ
U_M =  -\frac{1}{r^3 }\left[3(\MM_1 \cdot \er)(\MM_2 \cdot \er ) - \MM_1 \cdot \MM_2 \right], \label{UM0}
\EN
where $\er$ is the unit vector joining the two grains, and $\MM_{1,2}$ are the dipole moments.
Because we have assumed that the dipole moments are aligned with
the external field along the pole of our coordinate system, and we have assumed that the relative magnetic permeabilities of the grains
are equal, \Eq{UM0} reduces to
\EQ
U_M = \frac{\forall_{i1} \forall_{i2} M^2}{r^3} \left[1-3\cos^2(\theta)\right], \label{UM1}
\EN
where $(r,\theta)$ is the position of grain $2$, and $\theta=0$ is the pole.

We can calculate the radial magnetic force using \Eq{UM1}:
\EQ
\FF_M = - \partial_r U_M \er. \label{FM0}
\EN
Two point masses cannot interact in a way which would change their angular
momentum. Further, magnetic dipoles embedded in a uniform external magnetic field feel no net
force, but only a torque acting to align them with the external field. Accordingly, even though
$-\partial_\theta U_M \neq 0$, our system conserves its orbital angular momentum.
This can be understood by noting that while \Eq{UM1} is written in terms of $\theta$, the actual 
dependency in \Eq{UM0} is on the alignment of the magnetic dipole moments; and the apparent force $-\partial_\theta U_M$
actually represents the torques the two grains exert on each other to align their dipole moments: the torque drives spin, rather
than orbital angular momentum, and we are neglecting the spin of the particles. Because the particles would be torquing
each-other to stay magnetically
aligned, we would expect any spin they would develop to be aligned with, and extracted from, their orbital angular momentum.
Accordingly, our treatment of the motion of the grains as spin-free will underestimate the effect of their magnetic interactions.

As noted above, by assuming that the magnetization of the two grains is purely controlled by the external field, we are
neglecting the fact their magnetic dipole-dipole interactions will further work to align the dipole moments. Note that if the grains
are close enough together, the dipole moment of one can induce a dipole moment in the second, which
strongly alters the close-approach physics and further strengthens the attraction between the two particles \citep[see e.g.][]{SOFT_MAG}.
This simplification therefore results in a further underestimate
of the magnetic attraction for the two grains.

We can estimate the magnetic enhancement to the collisional cross-section by considering polar orbits (constant azimuth).
Using Equations~(\ref{UM1}) and (\ref{FM0}), the magnetic force on grain $2$ is
\EQ
F_M =  \frac{3\forall_{i1} \forall_{i2} M^2}{r^4 } \left[1-3\cos^2(\theta)\right]. \label{FM}
\EN
If grain $2$ has an initial velocity $v_\theta = v_0$ at an initial position $(r_0, \theta_0)$, then it has
angular momentum per unit mass
\EQ
L = r_0 v_0,
\EN
and hence a polar velocity
\EQ
v_\theta(r) = \frac{L}{r}. \label{vtheta}
\EN
We can use Equations~(\ref{FM}) and (\ref{vtheta}) to write
\begin{align}
&\frac{\partial^2 r}{\partial t^2} = \frac{F_M}{m_2} + \frac{v_{\theta}^2}{r} = \frac{3\forall_{i1} \forall_{i2} M^2}{m_2 r^4 } \left[1-3\cos^2(\theta)\right] + v_0^2 \frac{r_0^2}{r^3},
\label{Ftotal} \\
&\frac{\partial \theta}{\partial t} = \frac{v_\theta}{r} = \frac{v_0 r_0}{r^2}, \label{thetadot}
\end{align}
which define the motion of grain $2$.

Note that for $\theta_0=0$, the centrifugal and magnetic forces are balanced at a critical radius defined through
\EQ
r_c^3 \equiv  \frac{6\forall_{i1} \forall_{i2} M^2}{m_2 v_0^2 }, \label{rc}
\EN
which also lets us define the dynamical time scale
\EQ
t_c \equiv \frac{r_c}{v_0}. \label{tc}
\EN
Using Equations~(\ref{rc}) and (\ref{tc}) we can rewrite Equations~(\ref{Ftotal}) and (\ref{thetadot}) as
\begin{align}
&\frac{\partial^2 \tilde{r}}{\partial \tilde{t}^2} = \frac{1-3\cos^2(\theta)}{2 \tilde{r}^4}+ \frac{\left(r_0/r_c\right)^2}{\tilde r^3},  \label{rtildedotdot} \\
&\frac{\partial \theta}{\partial \tilde t} = \frac{\left(r_0/r_c\right)}{\tilde r^2}, \label{thetatildedot}
\end{align}
where $\tilde r \equiv r/r_c$ and $\tilde t \equiv t/t_c$.


We can integrate Equations~(\ref{rtildedotdot}) and (\ref{thetatildedot}) forward in time for given values of $r_0/r_c$ and $\theta_0$, determining
whether $r$ will hit zero (i.e.~a magnetic collision).
However, the limiting value of $r_0/r_c$ for which collisions occur (when one exists) for any given value of $\theta_0$ is complicated.
Nonetheless, the only control parameter for Equations~(\ref{rtildedotdot}) and (\ref{thetatildedot}) is $r_0/r_c$, so the
collisional cross-section for magnetic interactions between the two dust grains must scale with $r_c^2$. We can therefore write
\EQ
\sigma_m(v) = C \pi r_c^2 \label{sigma_m}
\EN
for the magnetic collisional cross-section, where $C$ is a constant. For $\theta_0=0$, $r_0/r_c=0.49$ is the limiting value for which
collisions occur, and we adopt $\bar{C} = 0.2 \lesssim .0.49^2$ as an approximation for $C$, noting that our analysis underestimates
the strength of the magnetic interactions, and that grains only need to approach to $r=a_1+a_2 \simeq a_1$ to collide.

This allows us to define the ratio of the magnetic collisional cross-section $\sigma_m $
 to the geometrical cross-section $\sigma_g \equiv \pi a_1^2$
 (recall that $a_1 \gg a_2$):
 \EQ
 \frac{\sigma_m(v)}{\sigma_g} =C\left(\frac{r_c}{a_1}\right)^2 = C \left[\frac{4 f_1 f_2 \phi \rho_s \mu_r^2 \rho_g k_B T}{\beta\rho_i^2\bar{m}   v^2 }\right]^{2/3}. \label{sigma_rat}
 \EN
 For velocities where $\sigma_m(v) \gg \sigma_g$, the magnetic cross-section applies, and when $\sigma_m (v) \ll \sigma_g$, the velocities
 are too high for magnetic forces to play a role, and the geometric cross-section applies. We approximate the intermediate regime
 by assuming that $\sigma=\max(\sigma_m, \sigma_g)$.
 
 Using \Eq{sigma_rat} we can define the critical magnetic velocity $v_m$ such that $\sigma_m(v_m) = \sigma_g$:
 \EQ
 v_m = 2 C^{3/4} \sqrt{ \frac{f_1 f_2\phi \rho_s \rho_g k_B T}{\beta\bar{m}}} \frac{ \mu_r}{\rho_i}. \label{vm_0}
 \EN
To calculate how much magnetic interactions change collision rates we will operate in terms of $v_m$, in which case \Eq{sigma_rat} becomes
simply
\EQ
\frac{\sigma_m(v)}{\sigma_g} = \left(\frac{v_m}{v}\right)^{4/3}.
\EN
Using the parameters in \Tab{normalizations}, we can write \Eq{vm_0} as
\EQ
v_m \simeq \left(\frac{C}{\bar{C}}\right)^{4/3} \sqrt{\frac{f_1 f_2 \phi \rho_g T \bar{\beta}}{\bar{f}^2\bar{\phi} \bar{\rho}_g \bar{T} \beta}} \frac{\mu_r}{\bar{\mu}_r} \times 0.2 \text{ cm s}^{-1}. \label{vm_1}
\EN
This estimate for $v_m$ is respectably high compared to estimates for the velocity at which dust grains bounced rather than
stuck together in the solar nebula, and justifies the remainder of this paper \citep{2010A&A...513A..56G}. Note that
$v_m \propto \rho_g^{1/2}$, so the strength of magnetic dipole interactions will generally decrease with orbital position.

\subsection{Magnetically enhanced collisional cross section: magnetically hard case}

In the case of magnetically hard magnetic dipoles not embedded in an external magnetic field, we can approximate that
the magnetic torques act to keep the dipole moments  optimally aligned. In that limit, \Eq{UM1} simplifies to
$U_M=-2\forall_{i1} \forall_{i2} M^2/r^3$. In this case, a grain with initial velocity $v_\theta=v_0$ at $r_0=r_c$ would
be on an unstable circular orbit, and the corresponding estimate for $C$ is $C = 1$. This would be a reasonable approximation
for dust grains each with only a single magnetic domain, and for which \Eq{vm_1} would become
\EQ
v_m \simeq 0.2 \left(\frac{C}{\bar{C}}\right)^{4/3} \sqrt{\frac{f_1 f_2 \phi}{\bar{f}^2\bar{\phi}}} \frac{M}{\bar{M}}\frac{\text{cm}}{\text{s}}
\simeq  171\sqrt{\frac{f_1 f_2 \phi}{\bar{f}^2\bar{\phi}}} \times \frac{\text{cm}}{\text{s}}, \label{vm_sat}
\EN
where we have used Equations~(\ref{M_norm}) and (\ref{M_sat}). More extreme than \Eq{vm_1}, this value for $v_m$ is large enough
for magnetized interactions to dominate dust coagulation even at scales often associated with fragmentation \citep{2010A&A...513A..56G}.
Note that observations of interstellar dust imply that only a small fraction of interstellar iron could be in the form of such single-domain
magnetically saturated iron-nickel dust grains
however \citep{1999ApJ...512..740D}.

\section{Effect of magnetization on collision rates}

\subsection{Turbulent collisional velocity distribution}

Our current picture for the collisional coagulation phase of dust growth at and below the size associated with chondrule precursors
is one where macroscopic grains are stirred by both turbulence and Brownian motion.
The degree of gas-dust coupling is captured by the dust Stokes number
\EQ
\St \equiv \tau \Omega,
\EN
where $\tau$ is the dust drag time and $\Omega$ the local Keplerian frequency.
Near the midplane of a vertically isothermal disk in vertical hydrostatic equilibrium
we have
\EQ
\St \simeq \frac{\sqrt{2\pi} a \rho}{\Sigma_g} \label{Stokes_base}
\EN
where $a$ is a dust grain's radius, $\rho$ its density and $\Sigma_g$ the gas surface density, 
and in this paper, where relevant, we assume that $\St \ll 1$. 
In that limit, turbulence drives a relative velocity between dust grains which scales as \citep{1980A&A....85..316V,2003Icar..164..127C}:
\EQ
v_t \propto \sqrt{\alpha \St} c_s \label{vt_0}
\EN
where the strength of the turbulence is measured through
the Shakura-Sunyaev $\alpha$ parameter \citep{1973A&A....24..337S}. We can see
from Equations~(\ref{Stokes_base}) and (\ref{vt_0}) that larger dust grains collide more violently than smaller ones.

If the two dust grains are very different in size, as we assumed in \Sec{sec_mag_interactions},
the turbulently driven velocity distribution is quasi-Maxwellian \citep{2013MNRAS.432.1274H},
and we have
\EQ
n(v,v_0) = n_0 \left(\frac{v}{v_0}\right)^2 e^{-v^2/2v_0^2}, \label{n_dist}
\EN
where $n(v)$ is the number density of targets with relative velocity $|v|$ per unit relative velocity.
The velocity scale $v_0$ is controlled by \Eq{vt_0}.  We assume \Eq{n_dist} henceforth, noting that
it also applies in the Brownian motion limit appropriate for very small particles extremely well coupled
to the gas.

\subsection{Collision rates}

The rate at which a dust grain will collide with other grains at velocities between $v_1<v_2$ is
\EQ
R(v_1,v_2) \equiv \int_{v_1}^{v_2} dv\, n(v)\, \sigma(v)\, v, \label{rate_0}
\EN
where $n(v)$ is the number density of targets with relative velocity $|v|$ per unit relative velocity,
and $\sigma$ the effective collisional cross-section (which may depend on the relative velocity).
In the absence of magnetic effects, we have only the geometric cross-section, i.e.~$\sigma=\sigma_g$.
Combining Equations~(\ref{n_dist}) and (\ref{rate_0}) we find:
\begin{align}
R_g\left(\frac{v_1}{v_0},\frac{v_2}{v_0}\right) &= n_0 \sigma_g v_0^2 \int_{\frac{v_1}{v_0}}^{\frac{v_2}{v_0}}du\,  u^3 e^{-u^2/2}, \\
&=-n_0 \sigma_g v_0^2\left[\left(u^2+2\right)e^{-u^2/2}\right]_{v_1/v_0}^{v_2/v_0}, \label{Rg0}
\end{align}
where the subscript $g$ (geometric) is used for non-magnetically mediated cases and we have performed
the non-dimensionalising substitution $u= v/v_0$.
Accordingly, the total geometric collision rate is given by
\EQ
R_g=R_g(0,\infty)=2 n_0 \sigma_g v_0^2,
\EN
where one must recall that $n_0$ is a number density per unit velocity.
Note also that $R$ (and hence all its variants in this paper) is a function of everything that the distribution $n$
depends on.  This reduces to $R$ also depending on $v_0$ if \Eq{n_dist} is assumed.

Magnetic interactions only matter for velocities $v<v_m$. Requiring $v_2<v_m$ we can therefore write
the rate of magnetically mediated collisions as:
\begin{align}
R_m\left(\frac{v_1}{v_0},\frac{v_2}{v_0}\right)&=  \int_{v_1}^{v_2} dv\, n(v,v_0)\, \sigma_m(v)\, v, \nonumber \\
&=n_0 \sigma_g v_0^{2/3}v_m^{4/3} \int_{\frac{v_1}{v_0}}^{\frac{v_2}{v_0}} du\, u^{5/3} e^{-u^2/2} \nonumber \\
\begin{split}
&=-n_0 \sigma_g v_0^{2/3}v_m^{4/3}\left[ \frac{2^{\frac 13}}{3} \Gamma\left(\frac 13 , \frac{u^2}{2}\right) + \right. \label{Rm0}\\
&\qquad  \qquad \, \, \, \, \, \left. \phantom{\Gamma\left(\frac 13 , \frac{u^2}{2}\right)}
u^{\frac 23} e^{-u^2/2} \right]_{\frac{v_1}{v_0}}^{\frac{v_2}{v_0}},
\end{split}
\end{align}
where $\Gamma$ is the incomplete Gamma function.
The total increase in interactions due to magnetic effects is just
\EQ
\bar{R}_m \equiv R_m(0,v_m/v_0)-R_g(0,v_m/v_0),
\EN
and so the fractional increase is
\begin{align}
\begin{split}
\frac{\bar{R}_m}{R_{g}} &= \left(\frac{v_m}{v_0}\right)^{\frac 43}\left[\frac{2^{-\frac 23}}{3} \left[ \Gamma\left(\frac 13,0\right)
-\Gamma\left(\frac 13, \frac{v_m^2}{2 v_0^2}\right)\right]\right] \label{Rm_R} \\
&\qquad \qquad \qquad \quad   \phantom{\Gamma\left(\frac 13, \frac{v_m^2}{2 v_0^2}\right)}+e^{-v_m^2/2v_0^2}-1. 
\end{split}
\end{align}
The important control parameter is clearly the ratio $v_m/v_0$: if it is large, then magnetic effects are important for the bulk of the potential
collisional partners, while if it is small most potential collisional partners are moving too fast for the magnetic effects
to play a role.

\begin{figure}\begin{center}
\includegraphics[width=\columnwidth]{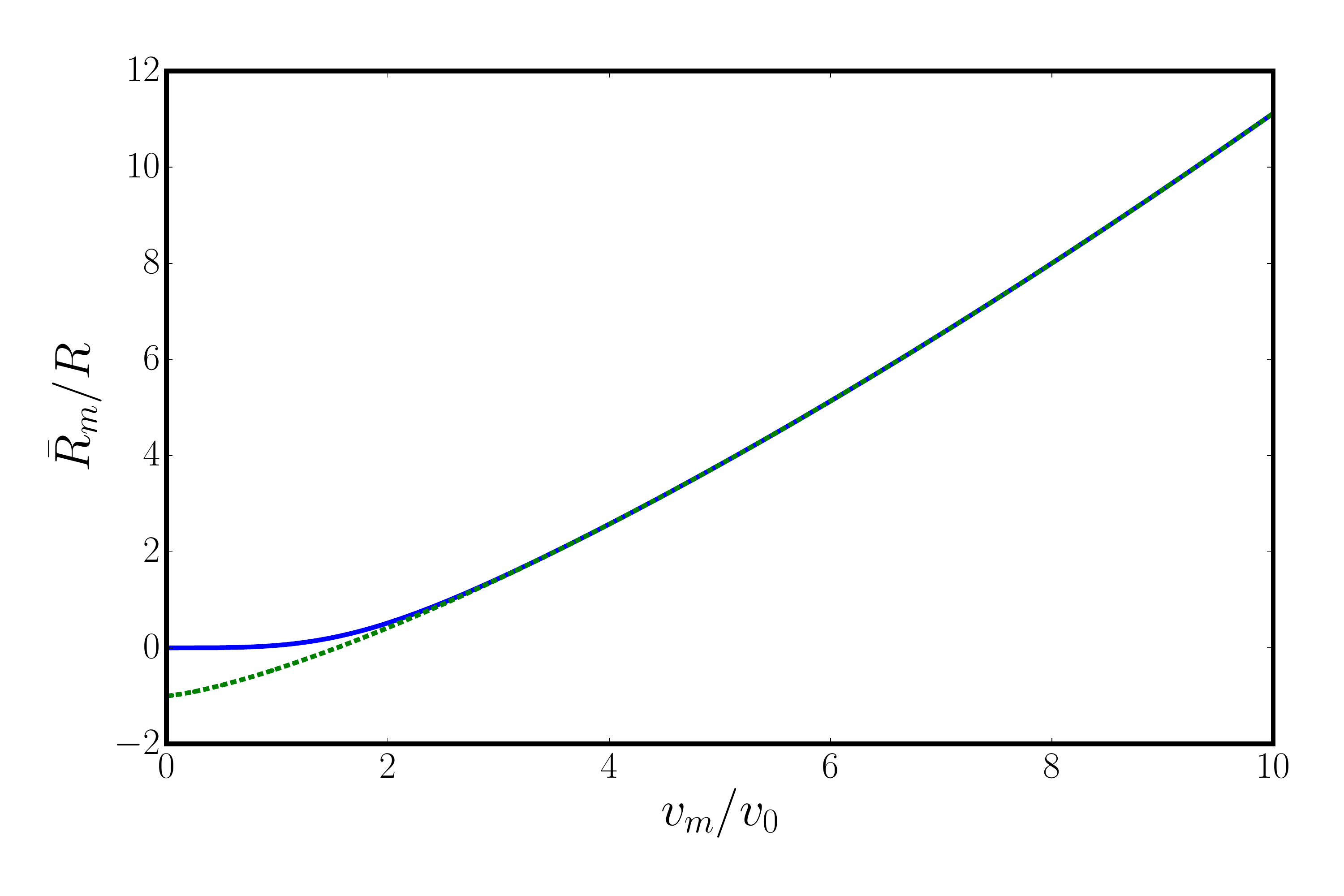}
\end{center}\caption{
Blue/solid curve: fractional increase in collision rate due to magnetic interactions as a function of $v_m/v_0$ (Equation~\ref{Rm_R}).
Green/dashed curve: large $v_m/v_0$ approximation (Equation~\ref{Rm_R_approx}).
\label{Rm_R_fig} }
\end{figure}

While \Eq{Rm_R} is somewhat complicated, many of its terms become negligible for large $v_m/v_0$, and we can approximate
\begin{align}
\frac{\bar{R}_m}{R_g}\left(\frac{v_m}{v_0} \gtrsim 2\right) &\simeq \frac{2^{-\frac 23}}{3} \Gamma\left(\frac 13,0\right)\left(\frac{v_m}{v_0}\right)^{\frac 43}-1 
\nonumber \\
&\simeq 0.56 \left(\frac{v_m}{v_0}\right)^{\frac 43}-1. \label{Rm_R_approx}
\end{align}
We plot both Equations~(\ref{Rm_R}) and (\ref{Rm_R_approx}) in \Fig{Rm_R_fig},
and one can see that the approximation is extremely good for $v_m/v_0 \gtrsim 2$. Further,
for $v_m/v_0 > 2.56$, $\bar{R}_m/R_g>1$ and magnetic interactions play a dominant role, more than doubling the collision rate.

\subsection{Varying $f$}

If we start with a population of dust grains with varying iron metal mass fractions, then $v_m \propto \sqrt{f_1 f_2}$ (see Equation~\ref{vm_0}).
For very small dust grains, which could even be to small to couple collisionally to turbulence and instead collide through Brownian motion
(which also has a Maxwellian distribution),
we expect $v_m \gg v_0$, and so $\bar{R}_m/R_g \simeq 0.56 (v_m/v_0)^{4/3}$ (further simplifying
Equation~\ref{Rm_R_approx}). In this case, the collision rate between iron rich grains
is enhanced compared to that between iron poor grains, which can help preserve an initial iron inhomogeneity. Take a scenario in which half
of the iron is found in metal poor grains making up $90\%$ of the population with
$f_{\text{mp}}= f_0$ (associated with a magnetic velocity $v_m$), while the other half is found in the $10\%$ of the grains that are metal rich
with $f_{\text{mr}}=9 f_0$. In that case,
the ratio of the rates at which iron rich grains will collide
with each other compared to colliding with iron poor grains is 
\EQ
\frac{0.1 \times (f_{\text{mr}}^2)^{2/3}}{0.9 \times (f_{\text{mp}} f_{\text{mr}})^{4/3}} \sim 0.48,
\EN
over four times the ratio that would occur in the absence of magnetic interactions. 
While collisional growth will still act to erase inhomogeneities,  magnetic interactions dramatically reduce
the rate at which it does so.

\section{Bouncing and fragmentation}

\subsection{Bouncing barrier}

Collisional dust growth is expected to
continue until the grains begin to collide
rapidly enough that electrostatic sticking forces are inadequate to hold them together upon impact and they begin to bounce. This will occur for collisional
velocities on the order of \citep{2010A&A...513A..56G}:
\EQ
v_b \sim 10^{-2} - 1 \text{\,cm\,s}^{-1}. \label{vb}
\EN
Those estimates are mostly below that of \Eq{vm_1}, so it is interested to explore the effects of magnetic interactions on
sticking rates. One must note that the release of magnetic potential energy means that magnetically mediated \emph{interactions}
occurring at $v=v_b$ will lead to \emph{collisions} at velocities $v>v_b$. However, 
this velocity increase is due to the magnetic potential energy well that the collisional participants
must exit for bouncing to occur; and we expect some of the kinetic energy to be thermalized or go into rearranging the dust grains. 
Magnetic interactions should therefore increase the effective bouncing velocity, although confirmation will
require future experimental and numerical work.
If the magnetically mediated collisions are violent enough, they could even lead
to fragmentation, although the iron components would remain bound \citep[e.g.~magnetic erosion,][]{2014arXiv1407.0274H}.
 For now, we assume that the critical velocities for 
bouncing are not changed by consideration of magnetic effects.

\begin{figure}\begin{center}
\includegraphics[width=\columnwidth]{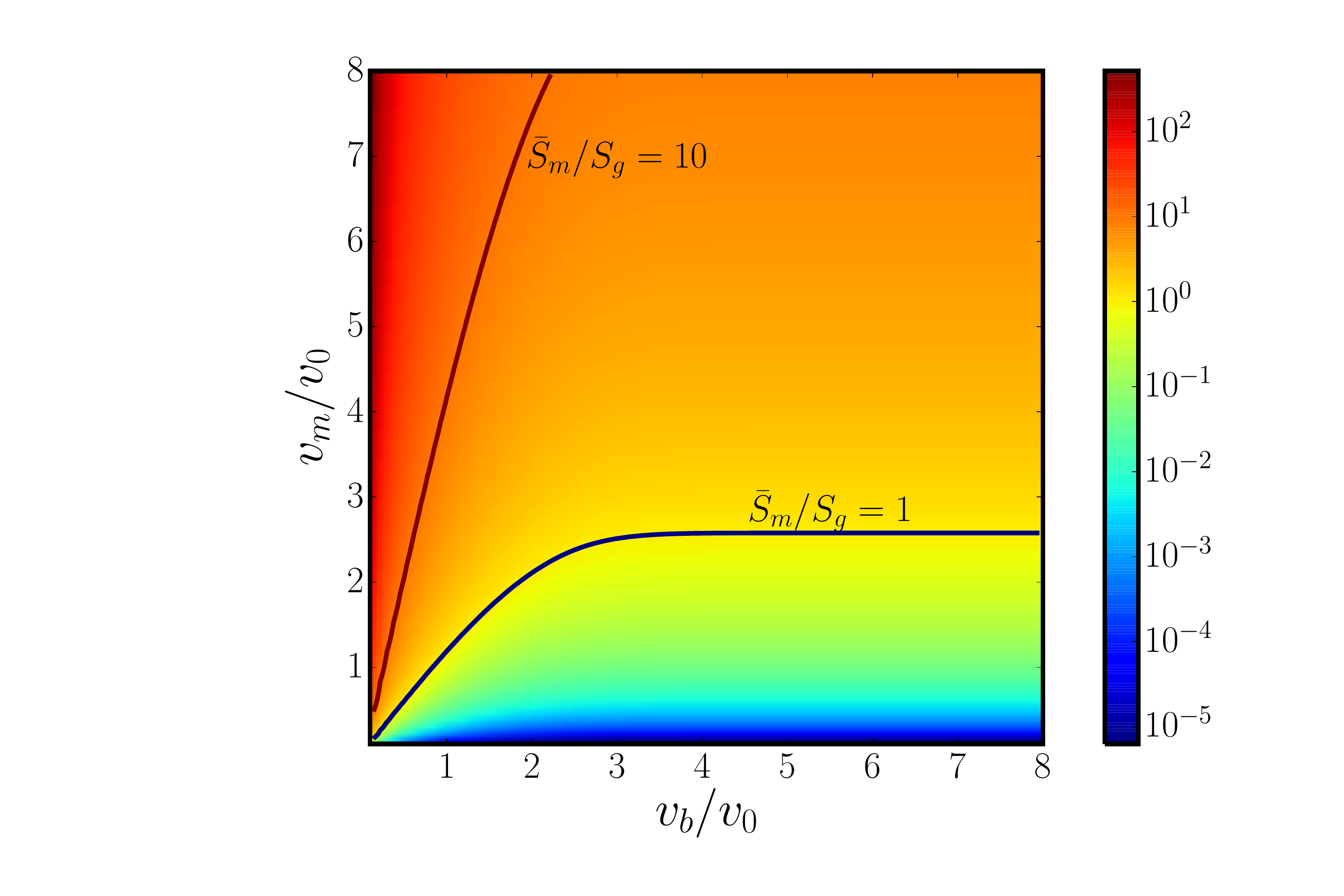}
\end{center}\caption{
Fractional increase in sticking rate due to magnetic interactions as a function of $v_m/v_0$ and $v_b/v_0$.
Top contour: magnetic interactions increase the sticking rate by a factor of $11$.
Bottom contour: magnetic interactions increase the sticking rate by a factor of $2$.
\label{Sm_S_fig} }
\end{figure}

Using \Eq{rate_0}, we can write the sticking rate as
\EQ
S \equiv R(0,v_b/v_0) = \int_{0}^{v_b} dv\, n(v,v_0)\, \sigma_{\text{eff}}(v)\, v. \label{rate_s_0}
\EN
In the absence of magnetic interactions we can use \Eq{Rg0} to find:
\EQ
S_{g} = 2n_0 \sigma_g v_0^2 \left( 1 - \left[1 + \frac{v_b^2}{2v_0^2}\right]e^{-v_b^2/2v_0^2}\right). \label{S_g}
\EN
Defining $v_c \equiv \min(v_b,v_m)$ and using
Equations~(\ref{Rm0}) and (\ref{S_g}) we can see that the increase in sticking rate due to magnetic interactions is given by
\begin{align}
\bar{S}_m &\equiv R_m(0,v_c/v_0)-R_g(0,v_c/v_0) \\
&=n \sigma_g v_0^2 \left( \frac{2^{\frac 13}}{3} \left(\frac{v_m}{v_0}\right)^{\frac 43} \left[ \Gamma\left(\frac 13,0\right)-\Gamma\left(\frac 13, \frac{v_c^2}{2v_0^2}\right)\right] 
\right. \nonumber \\
& \left. \qquad \qquad \quad  +\frac{2 v_0^2+v_c^2 - v_m^{\frac 43}v_c^{\frac 23}}{v_0^2} e^{-v_c^2/2v_0^2} -2 \right), \label{bS_m}
\end{align}
and the fractional increase is $\bar{S}_m/S_g$.
We plot $\bar{S}_m/S_g$ in \Fig{Sm_S_fig}. Note that for large $v_b/v_0 \gtrsim 3$ the ratio is approximately independent of $v_b$: almost all interactions
occur for velocities $v \sim v_0 \ll v_b$, so including bouncing has little effect. Note also that even for small $v_m/v_0$, magnetic interactions can
become arbitrarily important if $v_m>v_b$: all the interactions that don't bounce are magnetically mediated.

\subsubsection{Limiting approximations}

While the full form of $\bar{S}_m/S_g$ is lengthy, one is generally in the limit that $v_c \gg v_0$ or
$v_c \ll v_0$, and both limits admit simplifying approximations.
In the limit that $v_c \gg v_0$, we have (using Equation~\ref{Rm_R_approx}):
\EQ
\frac{\bar{S}_m}{S_g} \simeq \frac{\bar{R}_m}{R} \simeq 0.56 \left(\frac{v_m}{v_0}\right)^{\frac 43}-1. \label{Sm_S_approx}
\EN
This occurs because
only a negligible fraction of encounters are at a high enough velocity for bouncing to matter.

\begin{figure}\begin{center}
\includegraphics[width=\columnwidth]{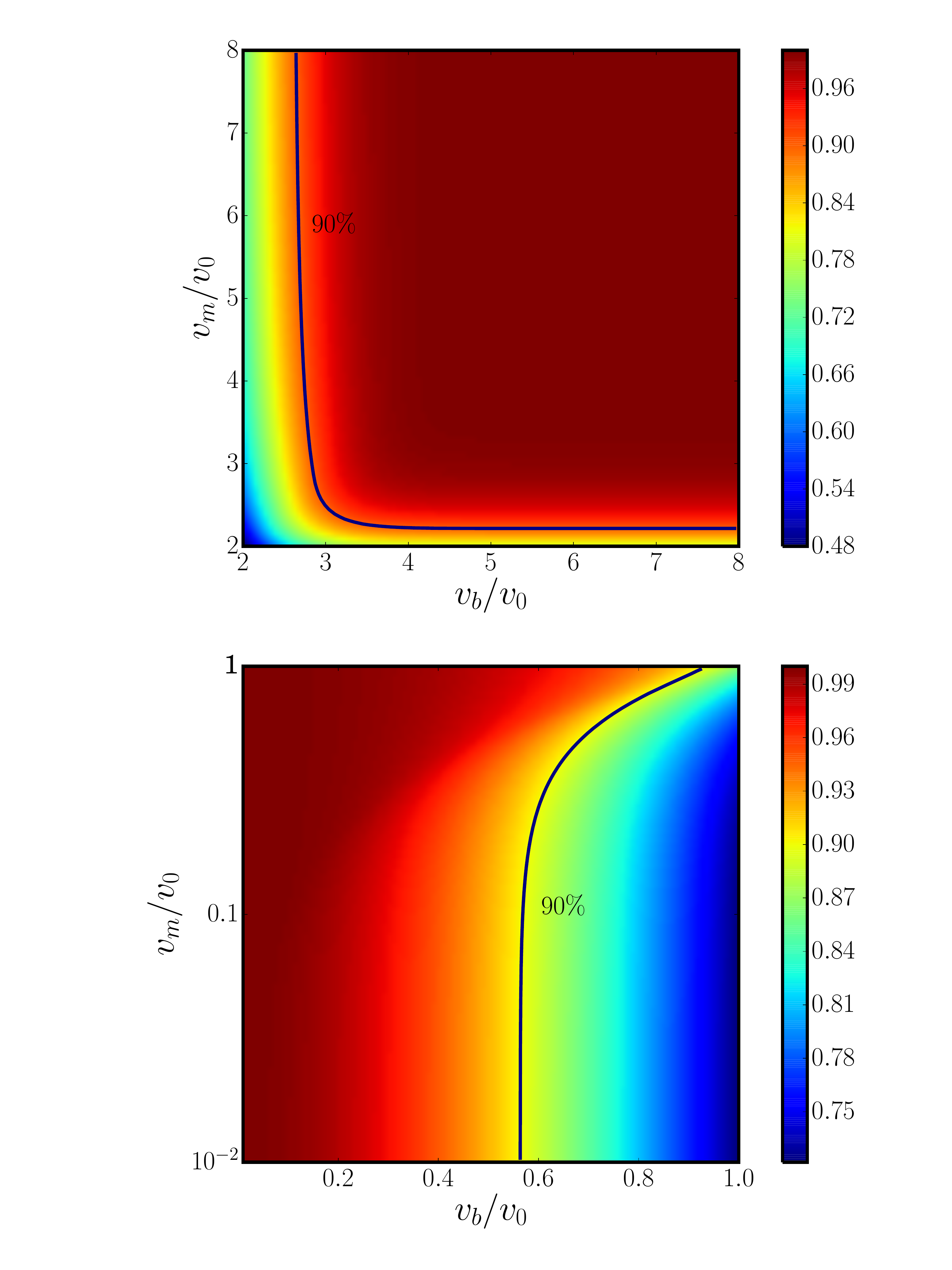}
\end{center}\caption{
Top panel:  Ratio of \Eq{Sm_S_approx}, the large $v_c$ approximation for $\bar{S}_m/S_g$, to the actual value. 
Bottom panel:  Ratio  of \Eq{Sm_S_low_approx}, the small $v_c$ approximation for $\bar{S}_m/S_g$, to the actual value. 
Contours are $90\%$.
\label{Sm_S_fig_approx} }
\end{figure}

In the limit of $v_c \ll v_0$, we can approximate \Eq{n_dist} as
\EQ
n \simeq n_0 \left(\frac{v}{v_0}\right)^2,
\EN
leading to
\begin{align}
&S_g \simeq  \int_0^{v_b} dv\, n_0  \sigma_g \left(\frac{v}{v_0}\right)^2 v = n_0 \sigma_g v_0^2 \times \frac 14 \left(\frac{v_b}{v_0}\right)^4,
\label{Sg_approx} \\
&R_g(0,v_c/v_0) \simeq n_0 \sigma_g v_0^2 \times \frac 14 \left(\frac{v_c}{v_0}\right)^4.
\end{align}
We also have
\begin{align}
R_m(0,v_c/v_0) &\simeq \int_0^{v_c} dv\, n_0 \sigma_m  \left(\frac{v}{v_0}\right)^2 v \nonumber\\
&\simeq n_0 \sigma_g v_0^2 \times \frac 38 \left(\frac{v_c}{v_0}\right)^{8/3}
\left(\frac{v_m}{v_0}\right)^{ 4/3}, \label{Rm_approx}
\end{align}
and hence
\EQ
\frac{\bar{S}_m}{S_g} \simeq \frac 32 \frac{v_c^{8/3}v_m^{4/3}}{v_b^4}-\left(\frac{v_c}{v_b}\right)^4. \label{Sm_S_low_approx}
\EN
We plot the ratio of these approximations for $\bar{S}_m/S_g$ to the actual value in \Fig{Sm_S_fig_approx}. We can see that
\Eq{Sm_S_approx} is a good approximation as long as we have both $v_m/v_0>2.5$ and $v_b/v_0>3$; and
that \Eq{Sm_S_low_approx} is a good approximation for $v_b/v_0< 0.5$.

\subsection{Balancing sticking and fragmentation}

In addition to bouncing, dust grains in protoplanetary disks colliding at velocities
\EQ
v > v_f \simeq 100\text{ cm s}^{-1} \label{v_frag}
\EN
will fragment \citep{2010A&A...513A..56G}. This value for $v_f$ lies well above our estimate for $v_m$ for magnetically
soft grains (Equations~\ref{vm_1}), but is comparable to the limiting value for saturated magnetically hard grains (Equation~\ref{vm_sat}).
Including  magnetic forces could change this picture by allowing magnetic erosion \citep{2014arXiv1407.0274H}, but exploring that
in detail is far beyond the scope of this paper; and will require future experimental or numerical studies of high velocity encounters between aggregates of monomers,
some of which are magnetized.  
Further, mass transfer is a possibility when colliding dust grains are very different in size, allowing for growth even at high collision
speeds \citep{2012A&A...544L..16W}.
For simplicity, we nonetheless  adopt the picture of \Eq{v_frag} for both geometric and magnetized cases.
Recall also that the collisional velocity scale $v_0$ is controlled by the dynamics of the disk. We have assumed turbulent stirring,
with $v_0$ scaling with the square root of the Stokes numbers of the dust grains (see Equations~\ref{Stokes_base} and \ref{vt_0}).

At some critical collisional velocity scale, dust grain collisions that result in sticking will balance those that result in fragmentation. We parameterize
this balance through
\EQ
R(0,v_b/v_0) = \psi R(v_f/v_0,\infty), \label{Eq_balance}
\EN
where $\psi$ is the critical growth-neutral ratio of the sticking rate to the fragmentation rate.
We also choose to define the ratio of the fragmentation and bouncing velocities as
\EQ
\xi \equiv v_f/v_b.
\EN
Note that $v_b$, $v_f$ and $\xi$ depend on the microphysics of the collisions, and we have assumed
that they are independent of any magnetic interactions. Combining Equations~(\ref{vm_1}), (\ref{vb}) and (\ref{v_frag}), we
expect $v_f \gg v_m > v_b$ for magnetically soft grains.
Accordingly, for a given dust grain size (and hence collisional velocity scale $v_0$), including
magnetic interactions would not effect the fragmentation rate, but would increase the sticking rate. If $\psi$ is constant, then
magnetized grains would find a balance between sticking and fragmentation for larger collisional velocity scales, and
hence at larger dust grain sizes. As is explored in \cite{Relative_Settling}, differences in dust grain sizes 
leads to aerodynamical sorting which could be correlated
with the chondrule formation process. Such a correlation would provide an explanation for the observed partitioning of iron,
and of tungsten isotopes, between chondrules and matrix \citep{1982GeCoA..46.1081G,2015LPI....46.2262B}.

In the geometric case we can consider \Eq{Eq_balance} to be an equation for
$v_b/v_0$ (and hence $v_f/v_0$) as a function of $\psi$ and $\xi$. In this case \Eq{Eq_balance}
becomes
\EQ
2-\left[\left(\frac{v_b}{v_0}\right)^2+2\right]e^{-\frac{v_b^2}{2v_0^2}}
= \psi
\left[\left(\frac{\xi v_b}{v_0}\right)^2+2\right] e^{-\frac{\xi^2 v_b^2}{2 v_0^2}}, \label{Eq_balance_g}
\EN
noting that $n_0$ and $\sigma$ have cancelled, and $v_0$ only shows up in ratio with $v_b$.
We write $u_g(\psi,\xi)$ as the solution of \Eq{Eq_balance_g} for $v_b/v_0$.
However, it is crucial to note that we have assumed that $v_b$ is set by the microphysics, so $u_g$ actually measures $v_0$.
Accordingly, we define
\EQ
v_{0,g}(\psi,\xi,v_b) \equiv v_b/u_g,
\EN 
the collisional velocity scale for \Eq{n_dist} at which sticking and fragmentation are balanced for
a set of parameters $(\psi, \xi, v_b)$.

For the magnetic case we can similarly use
\EQ
R_m(0,v_b/v_0) = \psi R_m(v_f/v_0,\infty) \label{Eq_balance_m}
\EN
to define $u_m(\psi, \xi, v_b, v_m)$ as the solution of \Eq{Eq_balance_m} for $v_b/v_0$.
We will use $u_m$ to define
\EQ
v_{0,m}(\psi, \xi, v_b, v_m) \equiv v_b/u_m.
\EN
The ratio $v_{0,m}/v_{0,g}$, which is also the ratio $u_g/u_m$ because $v_b$ and $v_f$ are not altered by magnetic interactions,
 describes how much more violently magnetized dust grains have to
be stirred for their collisions to be growth neutral; and hence how much larger they can grow.

\subsubsection{Exploring the consequences of magnetization on the balance between sticking and fragmentation}

The ratio $v_{0,m}/v_{0,g}$ depends on too many parameters to fully explore in this paper, but by restricting some
of the parameters we can estimate the impact that magnetization has on the dust grain size at which
sticking and fragmentation are balanced.
Based on Equations~(\ref{vm_1}), (\ref{vb}) and (\ref{v_frag}) we adopt
\EQ
v_b = 0.1 \text{ cm s}^{-1} < v_m \ll v_f = 100 \text{ cm s}^{-1}, \label{v_approxs}
\EN
i.e.~$\xi=10^3$. Given this ratio between $v_b$ and $v_f$ we can be certain that at least one of the limits $v_b \ll v_0$
or $v_f \gg v_0$ holds, and accordingly that at least one of the sticking or fragmentation rates is small. We will also
assume that $\psi$, the growth-neutral ratio of the sticking to fragmentation rates is neither  very large nor  very small.
Combining \Eq{v_approxs} with this condition on $\psi$ implies that both the sticking and fragmentation rates must be small,
and we have
\EQ
v_b \ll v_0 \ll v_f,
\EN
which allows significant simplifications.

\begin{figure}\begin{center}
\includegraphics[width=\columnwidth]{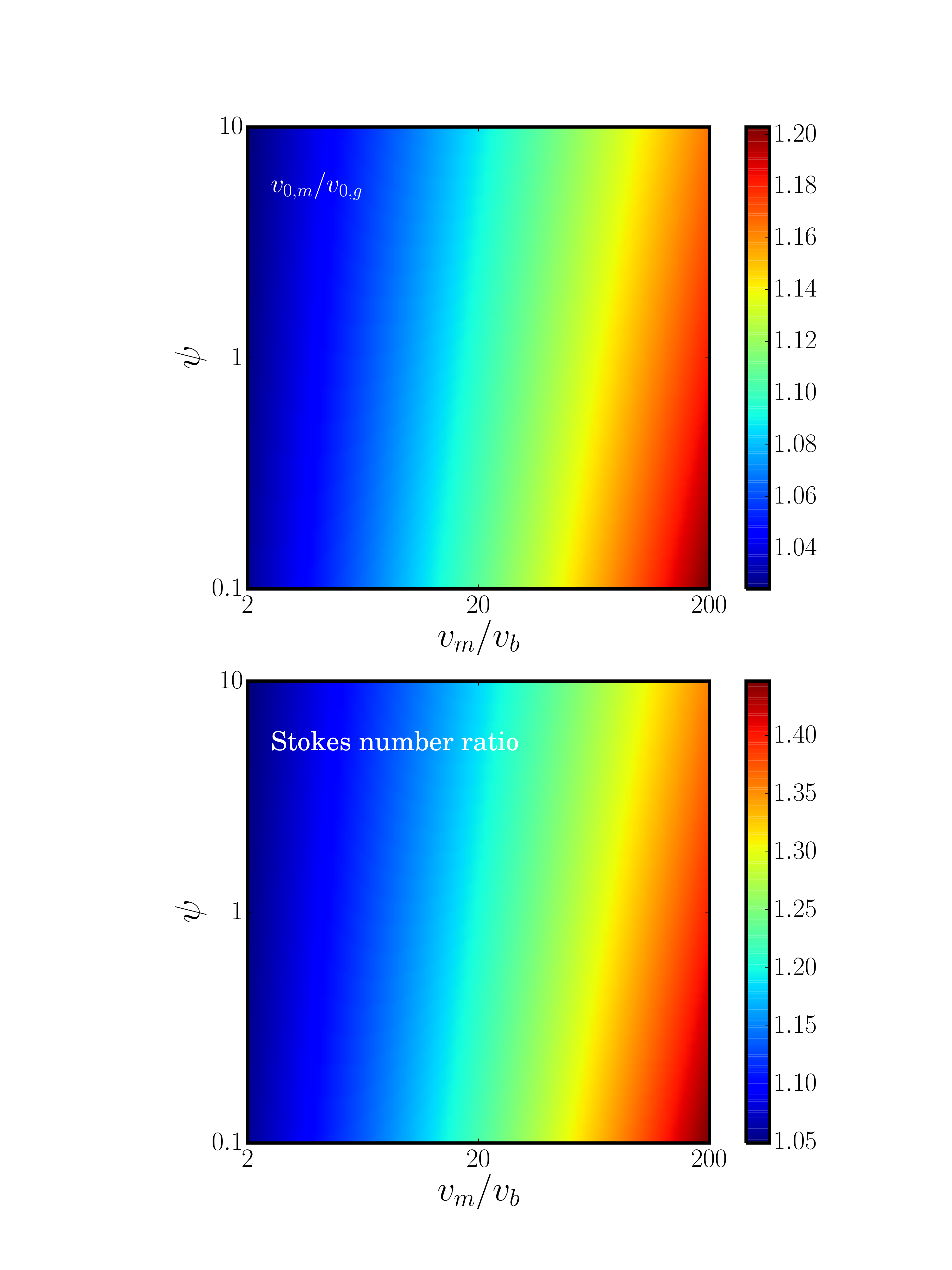}
\end{center}\caption{
Top panel:  Ratio of the turbulent collisional velocity scales required to balance sticking and fragmentation in the magnetized and non-magnetized cases
(the non-magnetized case does not depend on $v_m/v_b$). In this figure we hold $\xi \equiv v_f/v_b = 1000$ fixed, varying
the magnetic velocity $v_m$ and the growth-neutral sticking rate to fragmentation rate ratio $\psi$.
Bottom panel:  Ratio of the Stokes numbers of magnetically interacting growth-neutral grains to those of non-magnetically interacting
growth-neutral grains implied by top panel using Equation~\ref{vt_0}.
\label{fig_growth} }
\end{figure}

Because $v_b \ll v_0$ we can use \Eq{Sg_approx} to rewrite \Eq{Eq_balance_g} as
\EQ
4 \psi \left(\frac{\xi^2}{u_g^2} + \frac{2}{u_g^4}\right)e^{-\xi^2 u_g^2 /2} -1=0. \label{Eq_balance_g_0}
\EN
We also have $v_c \equiv \min(v_b, v_m) = v_b \ll v_0$ and $v_m\ll v_f$, so we can use \Eq{Rm_approx}
to rewrite \Eq{Eq_balance_m} as
\EQ
\frac 83 \psi \left(\xi^2 u_m^{-2/3} + 2 u_m^{-8/3}\right) e^{-\xi^2 u_m^2/2} - \left(\frac{v_m}{v_0}\right)^{4/3}=0. \label{Eq_balance_m_0}
\EN
We can parameterize the magnetic velocity $v_m$ in terms of the bouncing velocity and $u_m$ through:
\EQ
\frac{v_m}{v_0} = \frac{v_m v_{0,m} v_b}{v_{0,m} v_0 v_b} =  \frac{v_m}{v_b} \frac{v_{0,m}}{v_0}u_m. \label{vmv0}
\EN
Because \Eq{Eq_balance_m_0} is an equation for $v_{0,m}$ (through the intermediary $u_m$),
its solution for $v_{0,m}$ is unchanged by setting $v_0 = v_{0,m}$ in \Eq{vmv0}. This
reduces \Eq{Eq_balance_m_0} to
\EQ
\frac 83 \psi \left(\xi^2 u_m^{-\frac 23} + 2 u_m^{-\frac 83}\right) e^{-\frac{\xi^2 u_m^2}{2}} -
 \left( \frac{v_m}{v_b} u_m\right)^{\frac 43}=0. \label{Eq_balance_m_1}
\EN
For our fixed $\xi=10^3$, Equations~(\ref{Eq_balance_g_0}) and (\ref{Eq_balance_m_1}) 
can be  solved for $u_g$ and $u_m$ (and hence $v_{0,m}/v_{0,g}$) as functions of $\psi$ and $v_m/v_b$.
In \Fig{fig_growth} we plot the ratio $v_{0,m}/v_{0,g}$, along with the implied difference in Stokes number, for
growth-neutral grains. One immediate observation is that
the equilibrium size is only moderately shifted even when $v_m \gg v_b$ and magnetization
strongly increases the sticking rate.
We can understand this by taking the derivative of the fragmentation rate (i.e.~the right hand side of Equation~\ref{Eq_balance_g})
with respect to $v_0$:
\EQ
\partial_{v_0} \left(\psi
\left[\left(\frac{v_f}{v_0}\right)^2+2\right] e^{-\frac{v_f^2}{2 v_0^2}}\right)
\simeq \frac{\psi}{v_0} \left(\frac{v_f}{v_0}\right)^4 e^{-\frac{v_f^2}{2 v_0^2}},
\label{partial_v0_RHS}
\EN
where we have used $\xi v_b = v_f \gg v_0$. Dividing the fragmentation rate by its derivative we can estimate how sensitive
it is to small changes in $v_0$:
\EQ
\frac{\text{RHS}}{\partial_{v_0} \text{RHS}} \simeq v_0 \left(\frac{v_0}{v_f}\right)^2 \ll v_0. \label{sensitivity}
\EN
The exponential term in the fragmentation rate is an extremely sensitive function of velocity scale, 
so even significant
changes in the sticking rate will be balanced by small changes in the collisional velocity scale.

\subsubsection{Saturated magnetization case}

In the limiting case of magnetically hard grains magnetized to saturation (Equation~\ref{vm_sat}), we have
$v_m \sim v_f$, and magnetic interaction must in some fashion alter the critical collisional velocity.
We can estimate that if $v_m \gtrsim v_f$, where $v_f$ is the fragmentation speed in the absence of magnetic
interactions, then the magnetic interactions allow grain survival up to a magnetic-fragmentation speed
$v_{fm} \sim v_m$. In that case, we can use \Eq{vt_0} to estimate that the fragmentation limited dust grain size Stokes number
scales as
$\St \propto v_m^2 \propto f_1 f_2 \phi$.
In this limiting case, magnetic interactions are expected to strongly alter the maximum size to which dust grains grow.

\subsubsection{Varying f}

As noted above, while including magnetic interactions can decrease the time it takes for grains to grow enough for sticking 
and fragmentation to balance, the magnetic interactions do not strongly change the grain size at which that balance occurs.
Accordingly, iron rich and iron poor grains will have broadly similar sizes. Further, because we expect $v_m \ll v_f$, the magnetic energy
that helps hold iron rich grains together is much smaller than the kinetic energy associated with fragmentation.  Considerations
of magnetic interactions are unlikely to significantly change $v_f$.

Collisional fragmentation does not generally
completely destroy dust grains however, and it is possible that grains with inhomogeneously distributed iron
will fragment in a fashion that preserves the iron rich portions. Indeed, the flip side to \Eq{sensitivity} is that
almost all fragmentation collisions occur at velocities only a hair above $v_f$, do not have a significant energy excess available,
and must seek out the weakest links binding collisionally grown aggregates together. This could allow magnetic effects to play
an outsized role in controlling the results of the fragmentation process.

 If so, and given the more rapid regrowth of iron
rich dust grains, this could not only delay the loss of primordial iron metal inhomogeneities, but generate inhomogeneities
in an alternate version of magnetic erosion \citep{2014arXiv1407.0274H}.
We emphasis though that the difference in energy scales means that such a process would suffer from a fine-tuning problem.

\section{Conclusions}

We have examined the effect that magnetic dipole-dipole interactions would have on dust-dust collisions. The strength
of the interaction can be parameterized in a single velocity scale $v_m$, the critical velocity below which magnetic forces increase
the effective collisional cross-section. Under the approximation of a Maxwellian large-separation dust-dust relative velocity distribution,
appropriate for turbulently stirred particles of strongly differing sizes, or for Brownian motion,
the inclusion of magnetic interactions can be reasonably simply encapsulated.

In general, collision rates are strongly enhanced if the magnetic velocity $v_m>v_0$, the scale of the
Maxwellian velocity distribution. Further, if dust grain collisions
lead to sticking only below a characteristic velocity $v_b$, magnetic interactions strongly increase the sticking rate if either $v_m>v_0$ or $v_m>v_b$
is satisfied.
Finally, if collisions above a characteristic velocity $v_f$ lead to fragmentation, with $v_f > v_m$, then the inclusion of magnetic interactions
will increase the sticking rate, but not the fragmentation rate, increasing the size of growth-neutral dust grains whose sticking and fragmentation
rates are balanced. However, we find that this final effect is relatively minor due to the exquisite sensitivity of the fragmentation rate to
the Maxwellian velocity scale.

While the appropriate parameters are extremely uncertain, and especially in the case of the relative magnetic permeability $\mu_r$ could
vary by orders of magnitude, we have estimated for $v_m$ in the case of dust in protoplanetary disks whose iron
metal component is magnetized by the ambient magnetic field. We find that magnetic interactions
are expected to play a significant role in setting the dust-dust collision rates at least at and inwards of the orbital
position of our asteroid belt, at least for the smaller grains. This means that magnetic interactions must be considered
for the initial stages of dust coagulation, and dust chemistry, although the implications at the chondrule precursor scale are less certain
Indeed, if pre-solar iron was carried into the solar nebula in the form of single
magnetic domains and therefor maximally magnetized, magnetic interactions would play an overwhelming role. We hope that these
results inspire laboratory studies to explore the magnetic properties of plausible pre-solar iron-nickel bearing grains. Such
studies would not only provide better estimates for $\mu_r$, but would permit our model to be generalized beyond
perfectly magnetically soft and perfectly magnetically hard dust grains.

Because $v_m$ depends on the iron metal fraction
of the dust grains, magnetic forces will help preserve primordial iron metal inhomogeneities in the dust population of a protoplanetary disk,
providing an explanation for the observed partitioning of iron between chondrules and matrix \citep{1982GeCoA..46.1081G}.
 If those iron metal inhomogeneities were correlated with tungsten isotopic inhomogeneities, magnetic forces
could provide part of the puzzle for explaining the observations of \cite{2015LPI....46.2262B}. Both cases
also require that iron content be correlated with the chondrule formation process. We address this in a companion paper, \cite{Relative_Settling}, using a model
where the weak dependency of equilibrium grain size on magnetic interactions leads to iron rich grains being less coupled. In that case,
if chondrule formation events were limited to occurring at altitude, iron rich grains would be converted into chondrules
at a slower rate than iron poor grains.

Note however that the energy scale of fragmentation is much larger than that of the magnetic interactions in
the magnetically soft case, so a significant effect
at the largest dust size scale, where growth and fragmentation are balanced, is only expected if magnetic forces
help preserve iron rich portions of fragmenting dust grains. Most of the collisions above the critical fragmentation velocity $v_f$
are only just above that velocity though, and so are not expected to have a large energy excess. Therefore it is plausible that
magnetic effects could play an outsized role in controlling the fragmentation products, justifying future studies of high velocity impact
between collisional aggregrates, some of whose components are magnetized.

\section*{Acknowledgements}
This research was inspired and greatly helped by conversations with Denton S. Ebel, Harold Connolly and Michael K. Weisberg.
We thank the anonymous referee for pushing us to improve the sophistication of our analysis of the strength of the magnetic interactions.
The research leading to these results was funded by NASA OSS grant NNX14AJ56G.


\end{document}